\documentclass[acmtog, nonacm, balance=false,  table]{acmart}

\usepackage{amsmath}
\usepackage{booktabs} % For formal tables
\usepackage{float}
\usepackage{layouts}
\usepackage{wrapfig}
\usepackage{makecell}
\usepackage[ruled,vlined]{algorithm2e}
\usepackage{bbm}
\usepackage{appendix}
\definecolor{forestgreen}{rgb}{0.13, 0.55, 0.13}
\definecolor{lava}{rgb}{0.81, 0.06, 0.13}
\definecolor{magenta}{rgb}{0.7, 0.0, 1.0}
\definecolor{staticColor}{HTML}{005AB5}

\definecolor{dynamicColor}{HTML}{DC3220}

\newcommand{\reffig}[1]{Fig.~\ref{fig:#1}}
\newcommand{\refeq}[1]{Eq.~(\ref{eq:#1})}

\newcommand{\reftab}[1]{Table.~\ref{tab:#1}}
\newcommand{\refapp}[1]{App.~\ref{app:#1}}

\newcommand*{\img}[1]{%
    \raisebox{-0\baselineskip}{%
        \includegraphics[
        keepaspectratio,
        ]{#1}%
    }%
}

\newcommand*{\timestamp}[2][-0.5cm]{%
   \sffamily \small%
  \vspace{#1}%
  \begin{flushright}%
  \protect\img{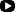}#2%
  \end{flushright}%
}

\usepackage{amsmath}
\DeclareMathOperator*{\argmax}{argmax}
\DeclareMathOperator*{\argmin}{argmin}

\usepackage{mfirstuc}
\MFUnocap{et}
\MFUnocap{al}

\newcommand{\W}{\boldsymbol{W}}
\newcommand{\B}{\boldsymbol{B}}
\newcommand{\0}{\boldsymbol{0}}

\newcommand{\y}{\boldsymbol{y}}
\newcommand{\Y}{\boldsymbol{Y}}
\newcommand{\G}{\boldsymbol{G}}

\newcommand{\q}{\boldsymbol{q}}

\newcommand{\D}{\boldsymbol{D}}
\providecommand{\d}{}
\renewcommand{\d}{\boldsymbol{d}}
\providecommand{\g}{}
\renewcommand{\g}{\boldsymbol{g}}

\newcommand{\f}{\boldsymbol{f}}

\newcommand{\Rn}[1]{\mathbb{R}^{#1}}

\newcommand{\F}{\boldsymbol{F}}
\newcommand{\R}{\boldsymbol{R}}

\providecommand{\a}{}
\renewcommand{\a}{\boldsymbol{a}}
\newcommand{\z}{\boldsymbol{z}}

\renewcommand{\r}{\boldsymbol{r}}
\newcommand{\Q}{\boldsymbol{Q}}
\newcommand{\J}{\boldsymbol{J}}
\newcommand{\A}{\boldsymbol{A}}

\newcommand{\x}{\boldsymbol{x}}
\newcommand{\X}{\boldsymbol{X}}

\renewcommand{\vv}{\boldsymbol{v}}
\providecommand{\vv}{}
\renewcommand{\b}{\boldsymbol{b}}

\providecommand{\u}{}
\renewcommand{\u}{\boldsymbol{u}}

\providecommand{\D}{}

\renewcommand{\D}{\boldsymbol{D}}
\providecommand{\H}{}
\renewcommand{\H}{\boldsymbol{H}}
\providecommand{\L}{}
\renewcommand{\L}{\boldsymbol{L}}
\providecommand{\M}{}
\renewcommand{\M}{\boldsymbol{M}}
\providecommand{\P}{}
\renewcommand{\P}{\boldsymbol{P}}
\providecommand{\I}{}
\renewcommand{\I}{\boldsymbol{I}}

\providecommand{\T}{}
\renewcommand{\T}{\boldsymbol{T}}

\newcommand{\bOmega}{\boldsymbol{\Omega}}

\providecommand{\N}{}
\renewcommand{\N}{\boldsymbol{N}}

\providecommand{\O}{}
\renewcommand{\O}{\boldsymbol{O}}

\providecommand{\V}{}
\renewcommand{\V}{\boldsymbol{V}}

\providecommand{\L}{}
\renewcommand{\L}{\boldsymbol{L}}

\acmJournal{TOG}
\begin{document}
% Title portion
% \title{Fast Complementary Dynamics}
% \title{Fast Complementary Dynamics via Skinning Subspaces}
\title{Actuators \`A La Mode: Modal Actuations for Soft Body Locomotion }
% DO NOT ENTER AUTHOR INFORMATION FOR ANONYMOUS TECHNICAL PAPER SUBMISSIONS TO SIGGRAPH 2019!

\author{Otman Benchekroun  }
\email{otman.benchekroun@mail.utoronto.edu}
\affiliation{
\institution{University of Toronto}
\country{Canada}
}
\affiliation{
\institution{Roblox Research}
\country{USA}
}
\author{Kaixiang Xie}
\email{}
\affiliation{
\institution{McGill University}
\country{Canada}
}
\affiliation{
\institution{Roblox Research}
\country{Canada}
}
\author{Hsueh-Ti Derek Liu}
\email{}
\affiliation{
\institution{Roblox Research}
\country{Canada}
}
\author{Eitan Grinspun}
\email{eitan@cs.toronto.edu}
\affiliation{
 \institution{University of Toronto}
 \country{Canada}
}
\author{Sheldon Andrews}
\email{}
\affiliation{
\institution{\'Ecole de Technologie Sup\'erieure}
\country{Canada}
}
\affiliation{
\institution{Roblox Research}
\country{USA}
}
\author{Victor Zordan}
\email{}
\affiliation{
\institution{Roblox Research}
\country{USA}
}
%\renewcommand\shortauthors{Zhou, G. et al}

%\begin{CCSXML}
\keywords{}

% <concept_id>10010147.10010371.10010352.10010379</concept_id>
% <concept_desc>Computing methodologies~Physical simulation</concept_desc>
% \ccsdesc[500]{Computing methodologies~Physical simulation}

\begin{abstract}
Traditional character animation specializes in characters with a rigidly articulated skeleton and a bipedal/quadripedal morphology. This assumption simplifies many aspects for designing physically based animations, like locomotion, but comes with the price of excluding characters of arbitrary deformable geometries.
To remedy this, our framework makes use of a spatio-temporal actuation subspace built off of the natural vibration modes of the character geometry.  
The resulting actuation is coupled to a reduced fast soft body simulation, allowing us to formulate a locomotion optimization problem that is tractable for a wide variety of high resolution deformable characters.
\end{abstract}

% uncomment for using teaser
\begin{teaserfigure}
   \centering%
  \includegraphics[width=\textwidth]{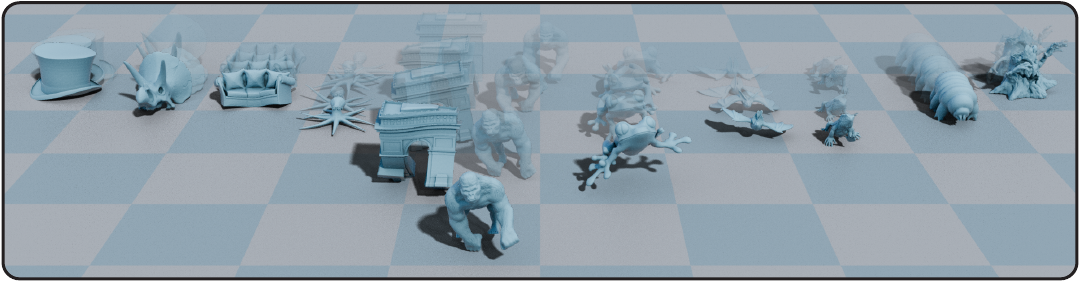}
  \caption{%
    We generate locomotion based off the actuations of the modal vibrations of a deformable character. We combine this actuation structure with a reduced order simulation to allow for the creation of locomotions for geometrically complex and varied soft-body characters. 
}
  \label{fig:teaser}
\end{teaserfigure}
\maketitle

\section{Introduction}
The world is comprised of a rich diversity of deformable organisms.
These organisms leverage their unique geometry and deformable nature for complex interactions with their environment in order to locomote, manipulate objects, or take flight.
They also serve as inspiration for the many fantastical creatures and virtual characters that populate video games, films, and virtual environments. 
%not to mention robots.  
% Therefore, having simulation and control methods that are able to faithfully capture the essence of these organisms and their behaviors is important for many applications in computer graphics and animation. 
%, and experience interacting with such virtual environments.
Having simulation and control methods that are able to faithfully capture the essence of these organisms and their behaviors is key to filling virtual worlds with these creative characters
However, most work on developing controllers for virtual characters is predicated on an extremely narrow set of morphologies that assumes an underlying rigid skeletal structure, and further imposes that they be bipedal or quadrupedal.

\begin{figure}[H]
    \centering
    \includegraphics[width=\linewidth]{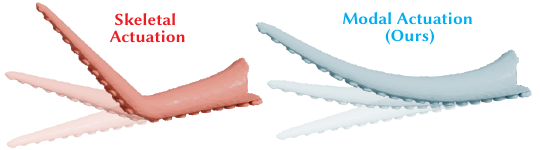}
    \caption{Traditional animation techniques an underlying skeleton structure, which is actuated via joint torques. The resulting actuation exhibits kinky, piecewise rigid deformation. Our actuation properly models the octopus tentacle as a deformable body, allowing for a naturally smooth deformation.}
    \label{fig:smooth_control_modes_vs_rigid}
\end{figure}

These assumptions preclude a wide range of characters, which are either bipedal or quadrupedal, from being animated.
% These assumptions preclude animating a wide range of characters %geometries 
% that do not fit the mold.
%rigid bipedal/quadripedal mold.
%
For instance, the motion of an octopus with tentacles as shown in \reffig{smooth_control_modes_vs_rigid} could not be accurately modeled using rigid segments, an issue that applies to a wide variety of deformable organisms.
%but it also affects the design of controllers for fantastical characters, whose geometry can be unpredictably exotic.
%
% This assumption places state of the art control methodologies out of reach for deformable characters.
% %
% This drawback is especially felt in fields like medicine, robotics, and biology, where soft deformable interactions can have hugely impactful positive or negative repercussions.
% %
% The sheer range of geometric qualities in the world around us has translated to a swath of geometric discretizations, such as point clouds, meshes, and implicit fields. 
% %
% This work aims to enable animating physics-based characters that are personified by soft and deformable behavior. 
%
These limitations motivate the need to develop locomotion controllers for soft-body characters with a wide range of highly detailed deformable mesh geometry, yet without the requirement of a piecewise rigid skeleton. 

Unfortunately, such controllers are difficult to generate for three reasons.
First, there exists \textit{significantly} less motion data for arbitrary geometries than there is for bipedal or quadripedal characters.
This data is key for creating natural-looking motions and its absence results in behaviors that exhibit unrealistic jittery motion \cite{heess2017emergence}. 
Second, deformable characters make use of a complex organization of various organic tissues, all interacting in harmony to carry out even the simplest of tasks \cite{keller2023skintoskeleton}. 
Designing this musculo-skelature for many characters becomes a tedious, unintuitive ordeal.

Finally, high resolution deformable characters usually come with the price of an expensive simulation, which scales in complexity with the resolution of the character geometry. 
%
% This scaling quickly makes creating a controller for a high-resolution deformable body an untractable problem.
% %
% Even simple controllers for rigid bodies typically require the evaluation of millions of simulation steps before convergence \cite{reda2020howenvironmentmattersforrl}.
% %
% This limits prior work on deformable characters controllers to extremely simple examples such as coarse geometries or 2D domains.

%
We propose a very simple, scalable pipeline for generating locomotion controllers for deformable characters of arbitrary user-defined geometry.  

The key to our method lies in defining a small data-free spatio-temporal actuation subspace based off the geometry of the character alone. 
An actuation signal in this subspace corresponds to a shape that the deformable character naturally wants to take on.
We then define a plasticity-based actuation energy, that guides our simulated character towards this actuated target shape.
This actuation is coupled to a \textit{reduced} deformable simulation, which allows us to perform the entire soft-body controller optimization within the reduced spaces of actuation and simulation, decoupled from the mesh resolution.
This allows us to achieve a variety of crawling, running and hopping-like locomotion behaviors with an off the shelf CMAES optimizer.%

We evaluate the effectiveness of our actuation subspace on a wide variety of character geometries, obtaining varied locomotions for highly detailed characters within minutes, while also providing avenues for intuitive motion control and design.
%
% We show we can achieve rich locomotion behaviors for a wide range of detailed meshed characters, while still providing avenues of intuitive motion control and design.

\section{Related Work}

\subsection{Soft Body Controllers}
Soft body characters bring with them many technical challenges in the design of locomotion controllers. 
%Creating locomotion controllers for soft-body characters has a number of technical challenges.
In particular, the physical model used for simulation of soft bodies is often characterized by a large number of degrees of freedom, bottle-necking the controller optimization.  

A limited amount of previous work addresses constructing soft body controllers, with the significant majority modeling the full space of the character geometry.  Due to the increased complexity from using high resolution models, most only  animate simple 2D or coarse 3D characters \cite{coros2012deformablebodiesalive, jain2011modal, tan2012softbodylocomotion}.
Even with simpler physical models and character geometries, 
\citet{bhatia2021evolution, lin2020softgym, huang2024dittogym,  hu2019difftaichi, rojas2021differentiable, min2019softcon} all report that training controllers for soft-body creatures with fewer than several thousand degrees of freedom can require hours, or even days, of training time, with simulations running much slower than real-time.

The large number of degrees of freedom (DOFs) required of soft-body simulations has motivated the development of reduced order models in order to improve simulation performance \cite{brandt2018hyperreducedpd, benchekroun2023fast, trusty2023subspacemfem}, with most of the prior work focusing on accelerating \textit{passive} simulation. 
ROMs have been leveraged to accelerate high resolution trajectory optimization tasks
\cite{pan2018reducedtrajectoryoptimization}, but their use remains limited for more general controllers.
While \citet{liang2023learningreducedordersoftrobotcontroller} accelerate a closed loop Model Predictive Control (MPC) controller, their approach still remains limited to simple and coarse character geometries, while exhibiting artifacts upon actuation \reffig{frog-actuation-comparison}.

\begingroup
\setlength{\abovecaptionskip}{-2pt} 
\begin{figure}[h]
    \centering
    \includegraphics[width=\linewidth]{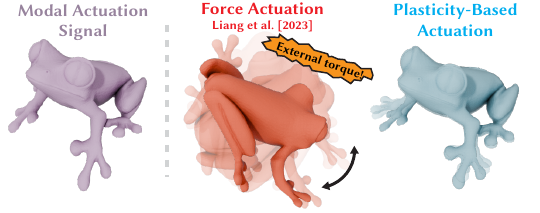}
    \timestamp{}
    \caption{Our plasticity based actuation energy conserves angular momentum, while taking on the expected target shape. The  force-based actuation from \citet{liang2023learningreducedordersoftrobotcontroller} does not conserve angular momentum, creating supernatural rotational motion upon actuation . }
    \label{fig:frog-actuation-comparison}
\end{figure}
\endgroup
\subsection{Designing Character Musculature}
Beyond fast simulation, an important step in designing controllers for a character lies in defining a viable musculoskelature for activating the physical system, and for arbitrary soft body characters, this is far from trivial.
By far the most common method of constructing this structure for rigid articulated characters is through the use of joint torques applied at the intersection of adjacent bones \citep{yin2007simbicon, 
 xu2023adaptnet}. These joint torques are either actuated directly, or indirectly through desired joint angles via PD controllers \cite{reda2020howenvironmentmattersforrl, tan2011stable}. 
Similar joint torques have been used to model soft-body characters whose motion is \textit{skeleton} driven \cite{libin2013skeletondrivensoftbodycontrol, kim2011direct, kim2011fast}, but this does not generalize to arbitrary soft body characters where a skeleton is either not obvious or non-existent.

Instead of joint torques, \citet{geijtenbeek2013flexible} use biomechanically inspired muscles, resembling springs with adjustable rest lengths, which ease the creation of contractible spring-like structures in deformable bodies \cite{rojas2021differentiable}.

Of course, the configuration and placement of muscles play an important contribution to the type of control that can be achieved.
\citet{lee2009biomechanicalmodellingupperbody, lee2014locomotion, lee2019scalable, saito2015computational, keller2023skintoskeleton} all propose highly detailed musculoskeletal models based on the real-world human anatomy.
However, as we explore different character morphologies, realistic anatomical models become ill-defined, and many methods rely on a user to manually design the placement of muscle fibers.
In the absence of an anatomical model, designers are left to their own devices, often opting to draw muscles directly onto the character \cite{tan2012softbodylocomotion, min2019softcon}. %themselves.

% can design three different types of muscles contractile muscles, radial muscles and . Then solves a trajectory optimization problem. Can draw contractive and bending muscles on character. Trains to actuate these muscles with a novel type of central pattern generator, modulated by DRL.

% \citet{liang2023learningreducedordersoftrobotcontroller} reduced simulation for simple characters, showing feedback control. The muscle structure is unspecified, and is instead learned by the controller itself. Expensive, but also unclear.
Alternatively, others shift the burden to an optimization procedure to design the control actuation structure \cite{lin2020softgym, bhatia2021evolution, ma2021diffaqua, liang2023learningreducedordersoftrobotcontroller}. 
The resulting optimization has a large search space, and has again been limited to coarse character geometries.

% Finally,  \citet{coros2012deformablebodiesalive} propose a muscle actuation structure governed by the pose of a coarse approximating cage, which updates the target rest shape of the character, parameterized by the low dimensional DOFs of the cage. 
% %
% This work closely matches our own, with the key difference being that our target pose is parameterized by the elastic eigenfunctions of the shape, rather than a user-defined coarse cage.

Yet another set of approaches from \citet{coros2012deformablebodiesalive, alecandru-eugen2017phace, pan2018reducedtrajectoryoptimization} propose a musculature defined by a plasticity-like deformation.
The character is encouraged to take on a changing target shape, as prescribed by a coarse cage, face scan data, or user-prescribed constraints and objectives.
Our approach similarly encourages the shape towards a new rest shape, but in our case, this rest shape is described by the natural vibrations of the character geometry.

\subsection{Motion Priors}
State-of-the-art character controllers often crucially make use of motion priors in order to guide the character to more natural behaviors. 
This is done by specifying the expected characteristics of the desired motion.

The copious amounts of motion data available for articulated characters lends itself well to this task, with many modern controllers making use of this data as reference motion that characters are encouraged to imitate  \cite{ peng2018deepmimic, peng2021amp,iccgan}. 
Aside from imitation, another effective method of leveraging this motion data is by constructing a motion subspace (or motion manifold \cite{starke2023deepphase}), either with PCA \cite{chai2005lowdimcontrolsignals} or more modern Deep Learning-based autoencoders \cite{ holden2015learningmotionmanifoldscnn, holden2020learnedmotionmatching, bergamin2019DreCon}.

Unfortunately, for soft body control of arbitrary geometries, motion data is not readily available. 
Thus, we must turn to alternative principles for guiding controllers, without the use of motion data.

For example, \citet{ranganath2021motorbabble} generate their own synthetic motion data for articulated characters, and derive an actuation subspace using Principal Component Analysis (PCA) applied to their synthetic data.
They show their deep reinforcement learning framework can animate articulated character with widely varying morphologies.

\citet{heess2017emergence} instead turn to a more rigorous training routine to fill the hole, observing that training a controller on a diverse set of environments leads to more robust rigid body controllers that better satisfy user defined objectives, albeit with some undesired jittery motion.

Periodicity has shown itself to be a particularly powerful prior, with
\citet{yu2018LearningSymmetricLocomotion} generating locomotion controllers that reward periodic and symmetric behaviors for articulated-bodies.
This can be taken one step further by imbuing periodicity into our controller actuation space itself;
Central Pattern Generators (CPG) \cite{guertin2009mammalian} model character motions with simple periodic control signals that are propagated across the character according to a low-dimensional set of parameters.
These kinds of temporal motion subspaces have been effective for generating 
 swimming and crawling motions \cite{min2019softcon, ma2021diffaqua, tu1994artificialfish,  tan2011articulated} and have been injected into modern deep reinforcement learning architectures to create rich encodings of human motions
\cite{holden2017phasefunctionnn, starke2023deepphase}

On top of periodicity, others also turn to energy efficiency to guide their animations. For example, \citet{kry2009modallocomotion} and \citet{ nunes2012naturalvibrationslocomotion} show that the natural vibrations of articulated characters can be used to generate kinematic locomotion animations for rigid characters of varying morphology.

Our approach extends these energy efficient, periodic subspaces demonstrated in prior works for soft body locomotion.
Where we derive an actuation subspace through the periodic actuation of the natural vibration modes of the deformable character.

\section{Method Overview}
Our goal is to derive an actuation for deformable characters that generates motion satisfying certain objectives, while remaining physically plausible.
We  cast this as a controller optimization problem,
\begin{align}
    &\min_{ \d(t)}\quad   J(\x(t)) \label{eq:objective_full_space}, \\
        \text{s.t.} \quad  \x|_t = &\argmin_{\x}   \quad   E(\x,  \d|_t) \quad \forall t  \in [t_0, t_1] \label{eq:full_space_sim},
        % \text{s.t.} \quad & \C(\x(t)) \leq \boldsymbol{0} 
    \end{align}
where we are solving for a time-varying actuation $\d(t) \in \Rn{3n}$ that corresponds to the target shape our character should take.
$J(\x(t))$ is a task-specific objective on the vertex positions $\x(t) \in \Rn{3n}$ that rewards desired motions (e.g for a locomotion task, it can reward motion of the center of mass along a target direction).

The constraint \refeq{full_space_sim} ensures the resulting motion abides the laws of soft-body physics with contact; The vertex positions at any point in time $\x|_t \in \Rn{3n}$ must be the minimizers of the total energy $E(\x|_t, \d|_t)$ of the physical system.
This physical energy can be split into two components,  
\begin{align}
E(\x, \d) = E_p(\x) + E_a(\x, \d),   
\end{align}
a passive component $E_p(\x)$ that captures motion according to inertia, external forces, and elastic forces, and an active component $E_a(\x, \d)$ that encourages the elastic body to take on the shape of $\d$.

On top of being non-linear in both $\x(t)$ and $\d(t)$, \refeq{objective_full_space} becomes especially difficult to solve for high resolution characters because the problem scales in complexity with the increased dimensionality of $\x(t)$ and $\d(t)$.
The sour combination of high non-linearity and high dimensionality makes solving this problem an extremely slow process prone to large null spaces.

Our solution to these problems is to introduce an actuation subspace for $\d$ based on the natural elastic vibration modes of the character \cite{pentlandwilliams1989goodvibrations}. 
%fOur 
We combine this reduction with a separate simulation subspace for $\x$, built off Skinning Eigenmodes \cite{benchekroun2023fast}. 
The separation of these two subspaces allows us to independently tailor each one to their own specific tasks, actuation and simulation, which have different desiderata.

We then show how to carry out simulation in these reduced spaces while remaining independent from the resolution of the mesh.
Our reduced-space framework allows us to quickly evaluate simulations of arbitrarily high resolution characters.

This speedup in simulation evaluation lends itself well to the use of a CMAES \cite{hansen2006cma} optimizer in order to solve our control optimization \refeq{objective_full_space}. 
This derivative-free genetic algorithm iteratively samples populations of $\d(t)$, which are used to drive multiple simulations. 
The resulting objective $J$ is measured for each simulation, and is used by CMAES to inform the collection of a next, more optimal generation of $\d(t)$.
The process is repeated until convergence, or until a maximum number of iterations are achieved.

\subsection{Modal Actuation Subspace}
\begingroup
\setlength{\abovecaptionskip}{-2pt} 
\begin{figure}
    \centering
    \includegraphics[width=\linewidth]{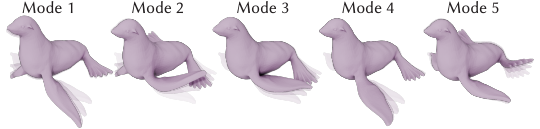}
    \timestamp{}
    \caption{The first 5 non-rigid elastic vibration modes of a seal, corresponding to reasonable low-energy motions one could expect to see from a seal.}
    \label{fig:seal-control-modes}
\end{figure}
\endgroup
Without access to motion data for arbitrary meshed characters, the guiding principles for defining our spatial-temporal actuation subspace become energy efficiency and periodicity.  

Drawing inspiration from modal analysis in computer animation~\cite{pentlandwilliams1989goodvibrations}, we build our subspace starting with the vibration modes of the character.
These are deformations of the character that induce the \textit{least} elastic energy, describing motions that the character \textit{prefers} to take as visualized on a seal in \reffig{seal-control-modes}.
While these vibration modes have been used by many prior works for accelerating passive simulations \cite{barbicjames2005realtimestvk, trusty2023subspacemfem}, we show that they form an effective subspace for soft body character actuation.

Vibration modes are the solutions to the generalized eigenvalue problem of the elastic energy Hessian $\H  \in \Rn{3n \times 3n}$,
\begin{align}
    \H \D = \M \D  \boldsymbol{\Lambda} \ ,
    \label{eq:motion-prior-spatial}
\end{align}
where the eigenvectors $\D \in \Rn{3n \times m}$ form an $m$-dimensional basis of \textit{spatial actuation modes}.
Above, $\M \in \Rn{3n \times 3n}$ is the diagonal vector-mass matrix and $\boldsymbol{\Lambda}$ is a diagonal matrix of eigenvalues. 

Deformed target positions $\d \in \Rn{3n}$ in this basis of actuation modes can be compactly represented by a low dimensional time-varying actuation vector $\a(t) \in \Rn{m}$  as well as the rest geometry $\x_0 \in \Rn{3n}$,
\begin{align}
    \d = \begin{bmatrix} \D \quad  \x_0 \end{bmatrix} 
    \begin{bmatrix}
        \a(t)  \\
        1
    \end{bmatrix} = \bar{\D} \bar{\a}(t).
    \label{eq:target-shape}
\end{align}
Where the $\bar{\mathrm{bar}}$ denotes a homogeneous expression of this actuation space $\bar{\D} = [\D \; \x_0] \in \Rn{3n \times (m +1)}$ and $\bar{\a} = [\a(t) \; 1]^T  \in \Rn{m+1}$, which we employ to simplify notation on future expressions.

With the intuition that most organisms make use of periodic motion patterns for locomotion, we imbue our actuation subspace with periodicity.
Our temporal actuation subspace is defined via a sum of $k$ sinusoids,
\begin{align}
    \a_i(t) =  \sum_j^{k} \boldsymbol{A}_{ij} \mathrm{sin}\left(2 \pi  \Big(\frac{t} {\T_{ij}} + \boldsymbol{\theta}_{ij} \Big)\right).
\end{align}
where $\A_{ij}, \T_{ij}, \boldsymbol{\theta}_{ij}$ are the amplitude, period and phase shift for each actuation mode $i$ and sinusoidal function $j$, which are parameters that are either learned by \refeq{objective_full_space}, or set by a user.

% \begin{figure}
%     \centering
%     \includegrathetacs[width=0.5\linewidth]{images/white.png}
%     \caption{Energy landscape as we vary the first control mode's amplitude, setting all other optimization parameters fixed. We show lines of different colors for varying the number of control modes with $m_s$, while keeping $m_t$ fixed to one. By decreasing $m_s$, the optimization problem becomes smoother and smoother.}
%     \label{fig:energy-landscape-reduction}
% \end{figure}

% \begin{figure}
%     \centering
%     \includegraphics[width=0.5\linewidth]{images/white.png}
%     \caption{ Many degrees of freedom can lead to an unnatural, jittery motion, restricting our optimization problem within our motion prior recovers smoother, more realistic gaits.}
%     \label{fig:jittery-motion}
% \end{figure}

%

% \input{04_method_new_2_skinning_eigenmodes_subspace}
\subsection{Reduced Simulation}
 With a low-dimensinoal actuation subspace in hand, we can actuate a soft body independently of its mesh resolution. 
 However, this resolution Independence does not transfer over to the simulation of the deformable, impeding forward simulation of high resolution characters.
 Worse yet, optimizing the parameters of the actuation subspace by solving \refeq{objective_full_space} requires many such forward simulations, becoming intractable as resolution increases.

For this reason, we also introduce a separate reduction of the soft body,
starting a linear Skinning Eigenmode Subspace \cite{benchekroun2023fast} for our vertex positions,
\begin{align}
 \x = \B \z   
\end{align}
where $\z \in \Rn{r}$ are the reduced space vertex positions and $\B \in \Rn{3n \times r}$ being our Skinning Eigenmode subspace matrix.  
We refer the reader to \refapp{skinning-eigenmode-subspace} for a refresher on the construction of this subspace.

With our two subspaces for simulation and actuation in hand, we must now rewrite the simulation optimization problem described in \refeq{full_space_sim} entirely in our reduced spaces.
We restate the full-space optimization problem in question, 
\begin{align}
   \x = \argmin_{\x}\; & E_p(\x) + E_a(\x, \d)   \nonumber 
\end{align}
with the goal of finding a method for solving this optimization problem, without ever touching the full resolution of the mesh.
Prior work \cite{barbicjames2005realtimestvk, jacobson2012fast,brandt2018hyperreducedpd, benchekroun2023fast} discusses extensively how to reduce passive soft body simulation and we refer the reader to \refapp{global-step-derivation} for details, summarizing that we largely follow the method of \citet{benchekroun2023fast} to arrive at a reduced passive approximation,
\begin{align}
    E_p(\x) \approx \tilde{E}_p(\z).
\end{align}
which can be evaluated entirely in a reduced space of  $r$ skinning eigenmodes and a set of $|\mathcal{C}_p|$ passive rotation clusters, as visualized in the appendix on a tardigrade \reffig{tardigrade}.

\subsubsection{Reducing the Actuation Energy $E_a(\x, \d)$}

We focus the remainder of our discussion on the actuation energy  $E_a(\x, \d)$.
This energy aims to make our simulated character $\x$ take on a target shape $\d$, and has the following form \cite{alecandru-eugen2017phace},
\begin{align}
\begin{split}
        E_a(\x, \d)  =    &\sum_e^{|\mathcal{T}|} 
    \min_{ \bOmega_e} \gamma_e V_e || \F_e(\x) - \bOmega_{e} \Y_e(\d) ||^2_F  \\
   &  s.t. \; \bOmega_e \in \mathcal{SO}(3).
    \label{eq:clustered-active-term}    
\end{split}
\end{align}
where $\F(\x)$ corresponds to the deformation Jacobian of the simulated character $\x$, while $\Y(\d)$ is the target deformation Jacobian of the actuation shape $\d$.
$\gamma_e$ and $V_e$ are the per-element actuation stiffness and volume respectively, the former of which can be tuned to achieve a stronger/weaker character. 
%

% Varying the stiffnesses $\mu_e$ and $\gamma_e$ allows us to control how much the character prefers to stay at rest instead of taking on the target shape $\d$, leading to different possible actuations as shown on a bat in 
% \reffig{bat-modal-actuation}
% \begin{figure}
%     \centering
%     \includegraphics[width=\linewidth]{images/bat_modal_actuation.pdf}
%     \timestamp{\tsBatActuation}
%     \caption{The actuation energy attempts to match a target modal displacement, even if that displacement incurs unnatural volume gains. \eitan{we've discussed this, but as a reviewer, this feels wrong.... i would want to use the local per-tet shape matching with no clustering, which wouldn't do this bad volume gain... i'd want to be convinced that the yucky volume gain is worth it, rather than passing the buck to the passive elastic term... you mentioned that the argument is that the single cluster makes the actuator faster at achieving the shape, compared to no clustering; but as a reviewer i might say what about no clustering and a much higher stiffness $\gamma_e$ for the active term?} The passive component of our actuation energy penalizes excessive deformation, and mitigates this issue. Their combination allows the bat to naturally flap its wings. }
%     \label{fig:bat-modal-actuation}
% \end{figure}

%
The matrix $\bOmega_e$ is a locally defined best fit rotation matrix, essential for filtering out rotations from our actuation;
an actuation signal $\d$ should lead to the same motion regardless of how the simulated element shape $\F_e(\x)$, and the target element shape $\Y_e(\d)$ are oriented.

This also imposes that our actuator conserves angular momentum, ensuring the character cannot supernaturally create an external torque.
This is not the case for a simple force-based actuation \cite{liang2023learningreducedordersoftrobotcontroller} which can introduce spurious linear and angular forces as shown in \reffig{frog-actuation-comparison}.

Plugging our subspaces for simulated positions $\x = \B \z$ and actuation target $\d = \D \a$ into this energy, 
\begin{align}
\begin{split}
        E_a(\z, \a)  =   \frac{1}{2} &\sum_e^{|\mathcal{T}|} 
    \min_{ \bOmega_e} \gamma_e V_e || \F_e(\z) - \bOmega_{e} \Y_e(\a) ||^2_F,  \\
   &  s.t. \; \bOmega_e \in \mathcal{SO}(3),
    \label{eq:clustered-active-term}    
\end{split}
\end{align}
leads us to a disappointing result; 
evaluating this energy requires the computation of the per-element best fit rotation matrix $\Omega_e$, requiring computation that scales with the number of tetrahedra in the character mesh.

\subsubsection{Clustering  $\; \bOmega_e$ for Mesh Independence of $E_a(\z, \a)$} 
To avoid recomputing $\Omega_e$ for each element in the mesh every simulation step, we derive a clustering scheme for our actuation energy, providing a reduction from the full number of elements $|\mathcal{T}|$.
Specifically, we allow for multiple elements to share the same rotation matrix $\bOmega_{c(e)}$, where $c(e) \in |\mathcal{C}_a|$ identifies the actuation cluster $c$, in the set of actuation clusters $|\mathcal{C}_a|$ to which an element $e$ belongs. 
Our approximation for the actuation energy becomes
\begin{align}
\begin{split}
         E_a(\x, \d) \approx  \tilde{E}_a(\x, \a) =  \frac{1}{2}\min_{ \bOmega_c}  \sum_e^{|\mathcal{T}|} 
    \gamma_e V_e || \F_e - \bOmega_{c(e)} \Y_e ||^2_F  \\
     s.t. \; \bOmega_{c(e)} \in \mathcal{SO}(3) \; \forall{c} \in |\mathcal{C}_a|.
    \label{eq:clustered-active-term}    
\end{split}
\end{align}

We prove in \refapp{optimal-clustered-rotations} that the optimal clustered rotation $\Omega_c$ minimizing the objective above can be found via a polar decomposition of the $\gamma V $-weighed sum of the covariance matrix $\F_e \Y_e^T$,
\begin{align}
    \bOmega_c  =   \mathrm{polar} \left( \sum_e^{|\mathcal{T}(c)|} \gamma_e V_e \F_e \Y_e^T \right) = \mathrm{polar} \left( \H_c \right) , 
    \label{eq:optimal_rotations}
\end{align}
where $\  \mathrm{polar}(\H_c)$ is a function performing polar decomposition on $\H_c$ in order to achieve its rotational component $\bOmega_{c(e)}$. 

While evaluating this term once again involves a per-element sum, we notice that this sum is a bilinear form in our two low-dimensional time-varying quantities, $\z$ and $\bar{\a}$. 
With this intuition, we show in \refapp{tensor-precomp} that by precomputing the appropriate tensor we can efficiently evaluate this sum at run-time in a manner scaling only in complexity with our small $\z$ and $\bar{\a}$,

\begin{align}
\H_c = \sum_{u}^r \sum_v^{m+1} \mathcal{H}_{cuv} z_u \bar{a}_v
% \H_{cij} &=    \mathcal{H}_{cijuv} z_u \bar{a}_v .
\label{eq:tensor_covariance_matrix}
\end{align}

Above, we have introduced the tensor $\mathcal{H} \in \Rn{|\mathcal{C}|  \times 3 \times 3 \times r \times(m + 1)}$, which we derive in \refapp{tensor-precomp}, that when multiplied against $\z$ and $\bar{\a}$,  computes our per-cluster covariance matrix $\H \in \Rn{|\mathcal{C}_a| \times 3 \times 3}$.
Note that evaluating $\H_c$ every stimulation step via this tensor product only requires operations that scale with the number of clusters $|\mathcal{C}_a|$, the dimension of our reduced positional DOFs $\z$, and the size of our actuation $\bar{\a}$.

As shown in \reffig{actuation-clusters}, a low number of actuation clusters allows for a more globally defined rotation invariance, with every character tetrahedron sharing the same rest frame.
This results in a simulated character that responds quickly to the actuation signal prescribed. 
Increasing the number of clusters actuatues each tetrahedron locally, independently of its neighbors, resulting in a slower response to actuation. 

%
% Ensuring these clustered rotation matrices remain the minimizers of the energy is crucial in preserving global rotation invariance, and in turn guarantees that the muscle forces are momentum conserving.

% To derive the clustered approximating energy for the passive muscle term $\tilde{\Psi}$, along with its efficient tensorial expression for $\R_{c(t)}$, one only needs to set  $\Y_e$ in \refeq{clustered-active-term} to the identity and swap $\gamma_e$ for $\mu_e$. This ends up giving the same clustered ARAP energy as \citet{jacobson2012fast}.
%
Finally, with our newfound ability to quickly evaluate our rotations, we arrive at the final form of our reduced and clustered simulation optimization problem.

\begin{align}
   \z = \argmin_{\z}\; & \tilde{E}_p(\z) + \tilde{E}_a(\z, \a)   
   \label{eq:global_step}
\end{align}

Where we have followed  prior work in reducing per-element non-linearities that appear in the passive term  \cite{jacobson2012fast}.

% \begin{align}
%               \argmin_{\z} \frac{1}{2h^2}|| \z - \q ||^2_{\N} + \tilde{\Psi}(\z) + \tilde{\Phi}(\z, \a),  
%               \label{eq:reduced-clustered-optimization-problem}
% \end{align}
% with $\q = \B \y \in \Rn{r} $ being our reduced inertial positions and $\N =\B^T \M \B  \in \Rn{r \times r}$  being the reduced space mass matrix. 
\begingroup
\setlength{\abovecaptionskip}{-2pt} 
\begin{figure}
    \centering
    \includegraphics[width=\linewidth]{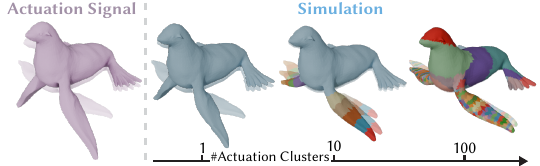}
    \caption{ Decreasing the number of actuation clusters results in a more globally actuated shape, accelerating the speed at which the simulation matches the target actuation signal. }
    \label{fig:actuation-clusters}
\end{figure}
\endgroup

\subsection{Local Global Solver}
\label{sec:local-global-solver}
We make use of a local global solver to solve our reduced simulation optimization problem.
This kind of solver comes with the advantage of maintaining a constant energy Hessian, allowing us to compute a factorization for this matrix once, then  reuse it throughout the entire rest of the optimization pipeline.
This is comprised of two main stages, a local step and a global step, that are repeated until a convergence criterion is met. 

The \textit{local} step first holds $\z$ fixed and optimizes for $\bOmega_{c}$ using \refeq{tensor_covariance_matrix} and \refeq{optimal_rotations}.
Then, the \textit{global} step holds $\bOmega_t$ fixed, optimizing for $\z$. This is resolved using a linear solve we derive in \refapp{global-step-derivation},
\begin{align}
    \Q & \z = \f ,
    \label{eq:local-global-system}
\end{align} 
with $\f \in \Rn{r} $ and  $\Q \in \Rn{r \times r} $ given by,
\begin{align}
    \Q = \Q^{p} + \Q^{a} \quad  & \quad 
    \f =  \f^{p} + \f^{a}  + \f^c. 
\end{align} 

The constant system matrix $\Q$ is a sum of the passive energy Hessian $\Q^{p}$ and our actuation energy Hessian $\Q^{a}$.
Details on the derivation of  $\Q^p$ and $\Q^a$ can be found in \refapp{global-step-derivation}.

The right hand side to the system in \refeq{local-global-system} $\f$ is a sum of the  passive  forces $f^p$,   actuation forces $f^{a}(\bOmega)$, and contact forces $\f^c(\z)$.
These contact forces are found in a projective fashion such that the resulting simulation after the global step \refeq{local-global-system} satisfies our no interpenetration, damping friction contact conditions.
Collision detection and handling is performed on a subset of the mesh vertices to maintain reduced complexity, as visualized on a tardigrade in the appendix \reffig{tardigrade}.
More details on the computation of this contact force can be found in \refapp{contact}.

\section{Results \& Discussion}
\label{sec:results}
% \begin{figure}
%     \centering
%     \includegraphics[width=\linewidth]{images/cma-ol-mesh-independance.pdf}
%     \timestamp{}
%     \caption{Varying the mesh resolution for the Arc de Triomphe mesh leads to a very similar locomotion style, with negligible extra optimization time.}
%     \label{fig:mesh-resolution-independance}
% \end{figure}

\begin{table*}[h]
\rowcolors{2}{white}{cyan!25}
\caption{ 
Locomotion optimization statistics for some of the locomotions shown in \reffig{teaser}. All these locomotions are found through 200 iterations of CMAES with a population size of 16, with a single actuation cluster. Below, \textbf{m} is the number of spatial actuation modes,  \textbf{k} is the number of temporal sinusoids used,   \textbf{r} is the number of skinning Eigenmodes,  \textbf{$|\mathcal{C}_p|$} the number of passive clusters, and  \textbf{$|\mathcal{I}|$} the number of contact samples.
\label{tab:locomotion-statistics}
}
\begin{tabular}{|c|c|c|c|c|c|c|c|c|c|c|}
\hline
\textbf{Mesh} & \textbf{\makecell{\#Vertices}} & \textbf{\makecell{\# Tets}}  & \textbf{m } & \textbf{k} & \textbf{r } & \textbf{\makecell{$|\mathcal{C}_p|$}} & \textbf{\makecell{$|\mathcal{I}|$}}  & \textbf{\makecell{$\mathbf{J}$}} & \textbf{\makecell{Sim. \\ Steps}} & \textbf{\makecell{Opt. \\ Time(s)}} \\
\hline
Couch & 13, 920& 64,154 & 10 & 2  & 5 & 5 & 20 & -0.63 &  300 &$9.84e^{2}$ \\
Creepy Tree & 42,015 & 200,487 & 6 & 2   & 5 & 5 & 40  & -0.39 &  300 & $3.3e^{3}$ \\
Bat & 9,266 & 34,720 & 3 & 2  & 5 & 10 & 20  & -1.25 &  300 &$9.00e^{2}$ \\
Bearded Dragon & 45,041 & 180, 406 & 6 & 2   & 7 & 10 & 30 & -0.97 &  200 &$1.27e^{3}$ \\
Octopus & 13,893 & 48,514 & 16 & 2  & 6 & 20 & 20 & -1.07 &  300 &$3.59e^3 $ \\
Treefrog & 13,771 & 54,154 & 5 & 3  & 5 & 10 & 30 & - 1.22 &  200 &$1.28e^3$ \\
% Add rows of data here
\hline
\end{tabular}
\end{table*}

% Our goal is to showcase the  rich variety of locomotions our actuation subspace can generate, with a very simple control optimization scheme.
% %
% With this in mind, we design a very simple locomotion objective, which we use to specify the control optimization problem \refeq{objective_full_space}.
We showcase the rich variety of locomotions our actuation subspace can generate by specifying the control optimization problem with a very simple objective \refeq{objective_full_space} 
\begin{align}
    \begin{split}
        J(\x(t)) &= J_{disp}(\x(t)) \cdot  J_{align}(\x(t)) \\
        J_{disp}(\z) &=  - (\x_{com}|_{t_1} - \x_{com}|_{t_0}) \cdot \hat{\vv} \\
        J_{align}(\x(t)) &= \min_t(\hat{\u}(t) \cdot \hat{\vv}),
    \end{split}
\end{align}
where $J_{disp}$ measures the distance travelled along a target direction $\hat{\vv}$ throughout the episode, and $J_{align}$  encourages the forward direction of the character, $\hat{\u}$ to point in that same target direction $\hat{\vv}$.
We solve the control optimization problem with open-loop controllers using an off-the-shelf implementation of CMAE-ES \cite{hansen2006cma}, using the $\texttt{pycma}$ library \cite{hansen2019pycma} and aim to directly predict the temporal actuation parameters, $\A$, and $\T$,  $\boldsymbol{\theta}$.

%  \begin{wrapfigure}[13]{r}{3.5cm}
% \includegraphics[width=3.8cm,keepaspectratio]{images/cma-ol-mesh-independance.pdf}
% \end{wrapfigure}

\paragraph{Controller Optimization Speed}
Our controller optimization is carried out entirely in a reduced space. 
This allows us to obtain locomotions for characters of arbitrarily high resolutions, with the slowest mesh we collect statistics on in \reftab{locomotion-statistics}, the octopus, taking just under an hour to locomote.
In contrast, using simulation statistics from a state of the art full space simulator of \citet{trusty2022mixed} on the smaller sized gecko mesh (a mesh smaller than all our examples),  and using the fastest locomotion optimization parameters we've tested (200 simulation steps, 200 CMAES iterations, population size of 16), would take an expected optimization time of at least 17 hours. 
Using this same analysis, all the examples in \reftab{locomotion-statistics}  are generously at least 17x faster than if we had used a full space simulation to optimize for their locomotion. 

\begingroup
\setlength{\abovecaptionskip}{-2pt} 
\newcommand*{\timestampmeshind}[2][-0.7cm]{%
   \sffamily \small%
  \vspace{#1}%
  \begin{flushright}%
  \protect\img{images/video-symbol.pdf}#2%
  \end{flushright}%
}

\begin{figure}
    \centering
    \includegraphics[width=\linewidth]{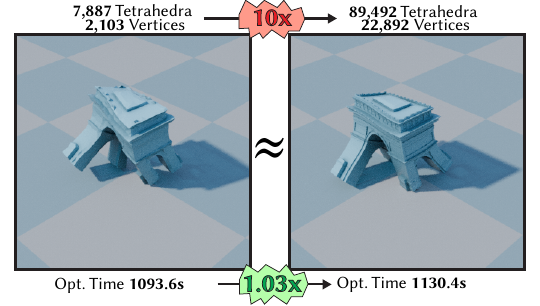}
    \timestampmeshind{}
    \caption{Varying the mesh resolution for the \textit{Arc de Triomphe} mesh leads to a very similar locomotion style, with negligible extra optimization time.}
    \label{fig:mesh-resolution-independance}
\end{figure}
\endgroup

\paragraph{Independence on Mesh Resolution}
Building further on our reduced space complexity,  \reffig{mesh-resolution-independance} shows the optimized locomotions of two different \textit{Arc de Triomphe} meshes: a visibly coarse one, and another that is finer by a factor of 10.
Despite large differences in resolutions the optimization time, as averaged over 5 different trials, is negligible between the two meshes.
In contrast, traditional full space local-global simulators \cite{bouaziz2014projective} have an $O(n^2)$ complexity to the resolution of the mesh, and an increase in resolution by a factor of 10 would result in a corresponding optimization time hit by a factor of 100.
Moreover, because our actuation subspace is defined via the elastic energy Hessian, which is convergent to the continuum of the geometry, we also get qualitatively similar locomotions across mesh resolutions.

\paragraph{Comparison to Different Actuation Subspaces}
\reffig{bat-cage-comparison} compares our actuation subspace with one defined by a coarse embedding cage, as proposed by \citet{coros2012deformablebodiesalive}.
A bat making use of these embedding cages can indeed find locomotions, however with a cage that is too low-dimensional, the locomotions make use of unnatural warping actuations, scaling/shearing the bat excessively.
Increasing the resolution of the cage to something that better conforms to the bat's geometry quickly increase dimension of the actuation subspace (for 32 cage vertices, the actuation subspace has 96 DOFs).
This increased dimensionality makes the bat prone to finding locomotions with small, shuffling steps, an observation echoed by \cite{pan2018reducedtrajectoryoptimization}.
In contrast, our modal actuations allow the bat to take more semantically meaningful deformations, such as flapping its wings, with only 3 actuation degrees of freedom.
Another factor that sets us apart from \citet{coros2012deformablebodiesalive} is our use of a reduced order physics model, which allows us to find locomotions for 3D examples of much higher resolution than theirs.

\begingroup
\setlength{\abovecaptionskip}{-2pt}
\begin{figure}
    \centering
    \includegraphics[width=\linewidth]{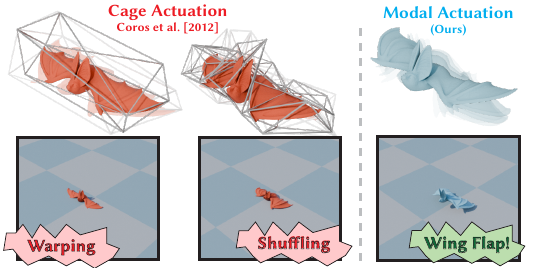}
    \timestamp{}
    \caption{Using a cage-based actuation introduces warping effects for low-resolution cages, or shuffling behaviors at high-resolution cages. Our modal actuation allows the bat to take more semantically intuitive motions to locomote, such as flapping its wings. }
    \label{fig:bat-cage-comparison}
\end{figure}

\paragraph{Exotic Actuation Modes}
Although we define our actuation modes through modal analysis, we also benefit from decades of work extending and generalizing modal analysis to obtain vibration modes with different qualities, such as sparsity \cite{brandt2017compressedvibration}, locality \cite{melzi2017localizedmanifoldharmonics}, discretization independence \cite{chang2023licrom} and non-linearity \cite{duenser2024compliantmodes, sharp2023data}.
Our method can leverage such advancements; We use the method of \citet{melzi2017localizedmanifoldharmonics} to design the actuation modes of the octopus to be locally bound to each individual tentacle. \reffig{smooth_control_modes_vs_rigid} shows such a local modal actuation, and \reffig{motion-style-control-freq}
shows the resulting locomotions these can generate.
\endgroup

\begingroup % Start of the group
\setlength{\columnsep}{10pt}
\setlength{\intextsep}{10pt}
\paragraph{Varying Actuation Subspace Dimension}
\begin{wrapfigure}{r}{0.5\linewidth}
  \includegraphics[]{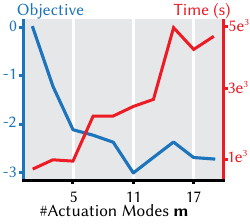}
\end{wrapfigure}
We motivate the importance of making use of a low dimensional actuation subspace in our optimization. 
We optimize a gorilla's locomotion parameters using an increasing number of spatial actuation modes, and track the resulting objective and total computation time until convergence.
The inset shows that an increased dimensionality for our spatial actuation prior is met with a plateauing of the character locomotion's objective, while requiring more optimization time until convergence.

\begingroup
\setlength{\abovecaptionskip}{-2pt} 
\begin{figure}
    \centering
    \includegraphics[width=\linewidth]{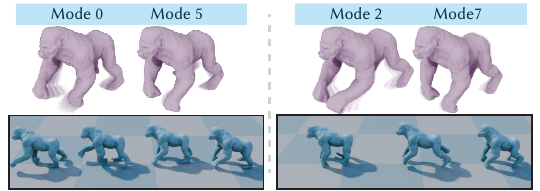}
    \timestamp{}
    \caption{ Using an actuation subspace composed of spatial modes moving the gorilla's limbs in sync (Left) vs. out of sync (Right). The synchronity of the limbs in the actuation subspace is reflected in the final optimized gait.}
    \label{fig:motion-style-control-modes}
\end{figure}
\endgroup

\paragraph{Motion Style Control}
Aside from allowing a user to select local bounds to each actuation mode, our actuation subspace provides other intuitive avenues for control.

By selecting which vibration modes to construct the actuation subspace, a user can generate different locomotion gaits.
For example, \reffig{motion-style-control-modes} shows a user picking different sets of actuation modes, one set that have the gorilla's limbs moving asynchronously, and another with limbs moving in synchrony. 
The resulting synchronicity of the limbs in the actuation subspace is intuitively reflected in the locomotion generated.

Yet another way to imbue user guidance into our framework is by having users constrain different parts of the temporal actuation parameters.
\reffig{motion-style-control-freq} shows a user specifying target periods our temporal sinusoids take on in order to generate locomotions varying in style.
Larger periods create more of a sluggish crawl for the octopus, while the smaller period makes the octopus skitter across the terrain with smaller steps.
\endgroup

\begin{figure}
    \centering
    \includegraphics[width=\linewidth]{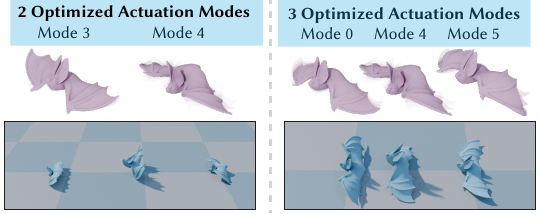}
    \timestamp{}
    \caption{ We can optimize for which control modes the bat should use to locomote, instead of selecting them arbitrarily. }
    \label{fig:co-optimizing-control-modes}
\end{figure}

\paragraph{Co-Optimizing for Actuations Modes}
While we have found selecting the first ~10 actuation modes as a default to usually generate viable locomotions, we can also ask our CMAES optimization to select \textit{which} subset of the top $m$ actuation modes to use. 
This requires a simple tweak in our optimization, where we add to our optimization degrees of freedom an actuation mode participation vector $\boldsymbol{\sigma} \in \Rn{m}$, whose highest valued entries chose the selected actuation modes used in the optimization.
Note that these can rapidly be queried and updated throughout the optimization without requiring any full-space matrix recomputations.
\reffig{co-optimizing-control-modes} shows the resulting locomotion for the bat, changing the number of desired queried control modes from 2 to 3.
While both provide locomotion, querying for three modes unexpectedly converges to a less optimal solution than querying only for two.
This happens due to the non-linear nature of our optimization problem with respect to the actuation parameters, making it susceptible to local minima.

% \begin{figure}
%     \centering
%     \includegraphics[width=\linewidth]{images/bat-autoselect-action.pdf}
%     \timestamp{}
%     \caption{ We can optimize for which control modes the bat should use to locomote, instead of selecting them arbitrarily. }
%     \label{fig:co-optimizing-control-modes}
% \end{figure}

% \begin{figure}
%     \centering
%     \includegraphics[width=\linewidth]{images/jumping.pdf}
%     \timestamp{}
%     \caption{Using only a single control mode to parameterize our motion prior, we can create an open loop jumping controller for the bat.}
%     \label{fig:jumping}
% \end{figure}

\begingroup
\setlength{\abovecaptionskip}{-2pt} 
\begin{figure}
    \centering
    \includegraphics[width=\linewidth]{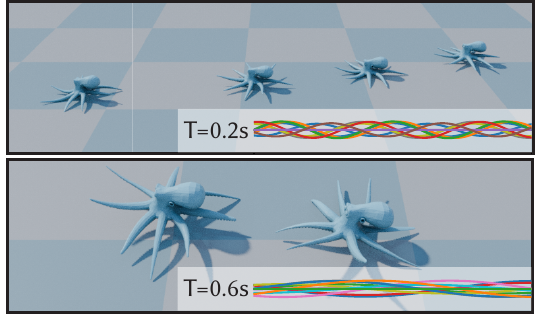}
    \timestamp{}
    \caption{A user can design faster-paced gaits by specifying the frequencies with which to actuate the actuation modes, and letting the optimizer identify the sinusoidal amplitudes $\boldsymbol{A}$ and phase shifts $\boldsymbol{\theta}$.}
    \label{fig:motion-style-control-freq}
\end{figure}
\endgroup

\begin{figure}
    \centering
    \includegraphics[width=\linewidth]{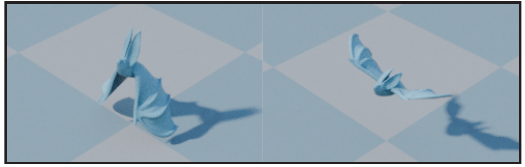}
    \timestamp{}
    \caption{Using only a single control mode to parameterize our motion prior, we can create an open loop jumping controller for the bat.}
    \label{fig:jumping}
\end{figure}

\paragraph{Jumping}
While our focus is on locomotion, \reffig{jumping} shows our actuation subspace can also be used to generate jumping motions for the bat. %
The only change required in our controller optimization lies in replacing the $J_{disp}$ objective with an objective that rewards mean center of mass height, $J_{height} = \mathrm{mean}(\x_{{com}_y})$.

\paragraph{Note on Realism}
The locomotions we obtain for many biologically inspired characters like the octopus are not \textit{realistic}, in the sense that a real living octopus wouldn't move forward by pushing against its back tentacles as shown in \reffig{motion-style-control-freq}.
This isn't surprising, our actuation structure by design has no concept of the real-world anatomy of an octopus.
Instead, our actuation forms a fictional\textit{virutal} musculature, generating motions that are 
\textit{geometrically} plausible, requiring little manual effort from the user to achieve a viable locomotion.

\paragraph{Robustness Across Geometries}
Finally, and most importantly, our method provides a framework for designing locomotions for characters of arbitrary geometry, without burdening the user with meticulous design of muscle fibers and bone connectivity \cite{tan2012softbodylocomotion, geijtenbeek2013flexible, min2019softcon}.
\reffig{teaser} shows locomotion on 11 very different high resolution soft body characters, each with their own unique actuation space and locomotion behavior.

\section{Conclusion and Future Work}
We have shown that the pairing of a periodic energy efficient actuation subspace with a reduced space soft body simulation provides a simple but powerful framework for designing locomotion for highly detailed characters with arbitrary deformable geometries.

%Limitations
We have many exciting avenues for future work.
First, our actuation subspace is built entirely off the elastic vibration modes of the character in isolation.
However, we believe a stronger actuation subspace could be built if we additionally made us of prior information regarding expected contact points and forces.
While our local-global solver is fast for softer characters, these solvers are known to converge slowly for highly stiff elastic material.
If one wanted to model locomotions for characters composed of extremely stiff materials, like steel and bones, 
we would recommend switching to other reduced solvers that are better equipped with dealing with such stiffnesses \cite{trusty2023subspacemfem}, which would come at the cost of abandoning our fast constant system factorization. 
%

% Future Work
While we have evaluated the effectiveness of our method on open loop controllers, many of the roadblocks we have cleared along the way are also bottlenecks in training deep reinforcement learning-based closed loop controllers.
We are especially excited about the possibility combining our framework to train such state of the art controllers, for more dynamics and controllable tasks and making soft-body characters replace articulated rigid ones as the default in the field of character animation.

\clearpage
\bibliographystyle{ACM-Reference-Format}
\bibliography{sample}

\clearpage
\newpage

\clearpage
\newpage
\appendix

\section{Skinning Eigenmode Positional Subspace}
\label{app:skinning-eigenmode-subspace}
We reduce our positional degrees of freedom $\x$ with a linear Skinning Eigenmode subspace \cite{benchekroun2023fast},
\begin{align}
 \x = \B(\W) \z   
\end{align}
where $\z \in \Rn{r}$ are the reduced space vertex positions. 

$\B(\W) \in \Rn{3n \times r}$ is a subspace matrix, parameterized by a set of Linear Blend Skinning weights $\W \in \Rn{w}$, with $r=12w$.
%
% We obtain these weights from a modal analysis on the elastic energy Laplacian, $\L = \frac{\partial^2 E}{\partial \x_x^2} + \frac{\partial^2 E}{\partial \x_y^2} + \frac{\partial^2 E}{\partial \x_z^2} \in \Rn{n \times n}$,
We obtain these weights from a modal analysis on the elastic energy Laplacian, $\L = \in \Rn{n \times n}$,

\begin{align}
\L \W &= \M \W \boldsymbol{\Lambda}.   
\label{eq:skinning_eigenmodes_gevp}
\end{align}

$\B$ is then constructed from $\W$ via the equation:
\begin{align}
\B &= \I_3 \otimes ( \W \otimes \boldsymbol{1}_4^T ) \odot (   \boldsymbol{1}_m^T  \otimes [\X \; \boldsymbol{1}_n] ),
\label{eq:lbs_jacobian}
\end{align}
Where $\X \in \Rn{n \times 3}$ is a matrix of stacked vertex positions at rest.

\section{Full Space Passive Energy}
\label{app:full-passive-energy}
The passive term is standardly comprised of a kinetic energy $E_k$,  an external potential energy for gravity $E_f$, and an elastic potential energy which we choose to be ARAP \cite{olgasorkine2007arap, kim2022dynamicdeformables} for all our examples $E_v$,
\begin{align}
     E_p(\x) = E_k(\x) + E_f(\x) + E_v(\x)
 \end{align}
\begin{align}
    \begin{split}
    E_k(\x) &= \frac{1}{2h^2} ||\x - \y ||^2_{\M},\\  
    E_f(\x) &= \x^T \M \g,\\
    E_v(\x) &= \frac{1}{2} \sum_e^{|\mathcal{T}|} \min_{\R_e} \mu_e V_e ||\F_e(\x) - \R_e ||^2_F \quad s.t. \; \R_e \in \mathcal{SO}(3) .
    \end{split}
    \label{full-passive-energy}
\end{align}

Above, $h$ is the simulation timestep size, $\y \in \Rn{3n}$ are inertial positions, $\g \in \Rn{3n}$ is the acceleration due to gravity acting on each vertex, $|\mathcal{T}|$ is the set of all mesh elements, $\mu_e$ and $V_e$ are the per-element elastic stiffness and volume respectively. $\F_e(\x)$ is the per-element deformation Jacobian of the simulated character, while $\R_e$ is its corresponding best fit rotation matrix.

\section{Reducing the Passive Energy}
\label{app:reducing-passive-energy}
We seek to detach the evaluation of the passive energy $E_p(\x)$, defined in \refapp{full-passive-energy},  with the resolution of the mesh. 
We follow an approach similar to \cite{benchekroun2023fast}, where we approximate our full space positions $\x = \B \z $, with the Skinning Eigenmode subspace from \refapp{skinning-eigenmode-subspace}.
To mitigate the evaluation of per-element rotations $\R_e$ in the elastic potential energy, we make use of the clustering scheme from \cite{jacobson2012fast}, where multiple elements may share the same rotation.
We call these clusters \textit{passive}, as they group together rotations from our passive elastic energy term. These are distinct from our \textit{active} clusters, which also allow for elements in a cluster to share the same rotation matrix.

Our active term finds the rotation matrix that aligns $\F_e(\z)$ to $\Y_e(\a)$, whereas the passive term finds the rotation matrix that aligns $\F_e(\z)$ to the identity. 
A distinction that can incur significant implementational differences in reduction.

Plugging in our active positional subspace and our rotation clustering scheme brings us to the reduced form of the passive elastic energy, which can be evaluated entirely in a reduced space.

\begin{align}
\begin{split}
 \tilde{E}_{k}( \z)   &= \frac{1}{2h^2}|| \z - \q ||^2_{\N} \\
   \tilde{E}_f(\z) &= \z^T \B^T\M \g,\\
\tilde{E}_v(\z) &= \frac{1}{2} \min_{\R_c} \sum_e^{|\mathcal{T}|}  \mu_e V_e ||\F_e(\z) - \R_{c(e)} ||^2_F \\ & \quad s.t. \; \R_c \in \mathcal{SO}(3)  \forall c \in |\mathcal{C}_p| . \\
 % \tilde{E}_{v}(\z, \a) &=  
 %    \frac{1}{2} \sum_e^{|\mathcal{T}|} \min_{\R_e} \mu_e V_e ||\F_e(\x) - \R_e ||^2_F \quad s.t. \; \R_e \in \mathcal{SO}(3) .
\end{split}
     \label{eq:contributing-terms-global-step}
\end{align}

\begin{figure}
    \centering
    \includegraphics[width=\linewidth]{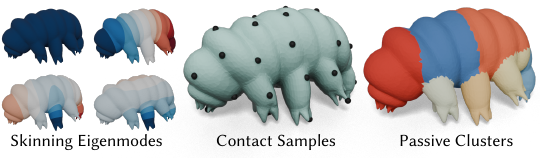}
    \caption{ We visualize the three simulation subspace parameters, namely 4 Skinning Eigenmodes, 10 passive muscle clusters and 30 contact sample points. }
    \label{fig:tardigrade}
\end{figure}

\begin{figure}
    \centering
    \includegraphics[width=\linewidth]{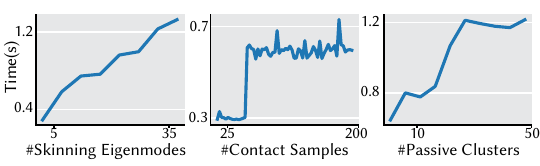}
    \caption{ Timings of 200 simulation steps as we vary three simulation subspace parameters of our reduced simulation, the number of skinning eigenmodes, contact samples, and passive muscle clusters. With the default for all quantities not being varied are 5 skinning eigenmodes, 40 contact samples, and 5 passive clusters. }
    \label{fig:timings}
\end{figure}

\reffig{tardigrade} visualize our Skinning Eigenmode Subspace and passive clusters on the tardigrade, while \reffig{timings} explores how thease two parameters affect simulation computation time.

\section{ Global Step Derivation for Local-Global Solver }
\label{app:global-step-derivation}
To solve for the character vertex locations in the reduced space $\z$, we make use of a local-global solver.
The local step optimize for a set of rotations both from our passive energy, and from our actuation energy, holding $\z$ fixed. These optimal rotations are derived in \refapp{optimal-clustered-rotations}. 
%
% The global step, we restate the entire reduced and clustered approximation of the physical simulation's optimization problem, 
The global step requires us to optimize for $\z$, while holding our rotation variables fixed.

We state our reduced space optimization problem

\begin{align}
   \z &= \argmin_{\z}\;  \tilde{E}_p(\z) + \tilde{E}_a(\z, \a)   \nonumber  \\
     &=   \tilde{E}_k(\z) + \tilde{E}_f(\z) + \tilde{E}_v(\z) + \tilde{E}_a(\z, \a) 
\end{align}
with all of our reduced space energies defined as,

\begin{align}
\begin{split}
 \tilde{E}_{k}( \z)   &= \frac{1}{2h^2}|| \z - \q ||^2_{\N} \\
   \tilde{E}_f(\z) &= \z^T \B^T\M \g,\\
\tilde{E}_v(\z) &= \frac{1}{2} \min_{\R_c} \sum_e^{|\mathcal{T}|}  \mu_e V_e ||\F_e(\z) - \R_{c(e)} ||^2_F \\ & \quad s.t. \; \R_c \in \mathcal{SO}(3)  \forall c \in |\mathcal{C}_p| . \\
\tilde{E}_a(\, \a) &= \frac{1}{2} \min_{\bOmega_c} \sum_e^{|\mathcal{T}|}  \mu_e V_e ||\F_e(\z) - \bOmega_{c(e)}  \Y_e(\a) ||^2_F \\ & \quad s.t. \; \bOmega_c \in \mathcal{SO}(3)  \forall c \in |\mathcal{C}_a| 
 % \tilde{E}_{v}(\z, \a) &=  
 %    \frac{1}{2} \sum_e^{|\mathcal{T}|} \min_{\R_e} \mu_e V_e ||\F_e(\x) - \R_e ||^2_F \quad s.t. \; \R_e \in \mathcal{SO}(3) .
\end{split}
     \label{eq:contributing-terms-global-step}
\end{align}

% \begin{align}
% \begin{split}
%  E_{k}( \z)   &= \frac{1}{2h^2}|| \z - \q ||^2_{\N} \\
%     \tilde{E}_{\Psi}(\z) &=  \sum_t^{|\mathcal{T}|}
%     \mu_t || \F_t - \R_{c(t)}  ||^2_F,  \\ 
%        \tilde{E}_{\Phi}(\z, \a) &=  \sum_t^{|\mathcal{T}|} 
%     \gamma_t || \F_t - \bOmega_{c(t)} \Y_t ||^2_F      
% \end{split}
%      \label{eq:contributing-terms-global-step}
% \end{align}

The global step assumes we've already found the minimizing rotations $\R_c$ and $\bOmega_c$ for the current iterate, and so we can treat these as constants and drop the two nested minimization problems for these two quantities, as well as the constraints that had appeared in the full energy \refeq{clustered-active-term}. 
%
% \derek{what is the constraint in \refeq{clustered-active-term} that get included here? It seems like the only constraint in \refeq{clustered-active-term} is just about $\bOmega_{c(e)} \in SO3$? If so, maybe just write it out that $\bOmega_{c(e)} \in SO3$ in the above equation. It takes less space than all these words about seeing \refeq{clustered-active-term} for constraints, and improves clarity.}

% \begin{align}
%    \tilde{E}_{\Psi}(\z) &= \sum_t^{|\mathcal{T}|}
%     \mu_t || \F_t - \R_{c(t)}  ||^2_F, \quad
%       \tilde{E}_{\Phi}(\z, \a) &=   \sum_t^{|\mathcal{T}|} 
%     \gamma_t || \F_t - \bOmega_{c(t)} \Y_t ||^2_F .
% \end{align}

Our strategy will be to expand all energies in \refeq{contributing-terms-global-step} and drop any terms that explicitly do not depend on $\z$, as they have no bearing on our optimization task.

\subsection{Expanding the  Elastic Energy $\tilde{E}_{v}(\z)$}\label{app:Ev_derivation}
Starting with the passive energy in \refeq{contributing-terms-global-step}, we expand out the Frobenius norm and obtain,
\begin{align}
  \tilde{E}_{v}(\z)  &= \frac{1}{2} \sum_e^{|\mathcal{T}|}
    \mu_e V_e || \F_e - \R_{c(e)}  ||^2_F \nonumber,  \\
          &= \sum_e^{|\mathcal{T}|}
    \frac{\mu_e V_e}{2}   tr(\F_e^T \F_e) -  \mu_e V_e tr(\F_e^T \R_{c(e)}) +  \mathrm{const}.
\end{align}
We deal with the first term  $ \mu_e V_e tr(\F_e^T \F_e)$, and rewrite the integrand inside the sum in index notation, reintroducing our definition for $\F_{eij} = {\mathcal{J}^{\B}}_{eijk}z_k $ from \refapp{tensor-precomp}.
\begin{align}
    % (\B \A)_{ij} =  B_{ik} A_{kj} \\
    % (\B^T \A)_{ij} =  B_{ki} A_{kj} \\
    % (\A \A)_{ij} =  A_{ik} A_{kj} \\
    % (\A^T \A)_{ij} = A_{ki} A_{kj} \\
    % tr( \C) =  C_{ii} \\
    % tr(\A^T \A) = (\A^T \A)_{ii} =   A_{ij} A_{ij} \\
    % tr(\F^T \F)  = (\F^T \F)_{ii} =   F_{ij} F_{ij} \\
    % \mu  tr(\F^T \F)  = \mu F_{ij} F_{ij} =  \mu {\mathcal{J}^{\B}}_{ijk} z_k {\mathcal{J}^{\B}}_{ijl} z_l \\
    \begin{split}
     \mu_e V_e  tr({\F_e}^T \F_e) &= \mu_e V_e ({\mathcal{J}^{\B}}_{eijk}z_k) ({\mathcal{J}^{\B}}_{eijl}) z_l \nonumber ,\\ 
     &= \mu_e V_e {\mathcal{J}^{\B}}_{eijk} {\mathcal{J}^{\B}}_{eijl} z_k z_l, \nonumber \\
      & =  Q^{v}_{ekl} z_k z_l,
    \end{split}
\end{align}
with $Q^{v}_{ekl} = \mu_e V_e {\mathcal{J}^{\B}}_{eijk} {\mathcal{J}^{\B}}_{eijl}$.
Plugging this back into the sum gives us
\begin{align}
      \sum_e^{|\mathcal{T}|} Q^{v}_{ekl} z_k z_l &=  z_k  (\sum_e^{|\mathcal{T}|}   Q^{v}_{ekl})  z_l = \z^T \Q^{v} \z,
\end{align}
where we've exposed the Laplacian matrix for our actuation energy $\Q^{v} = \B^T \J^T  (\V^{v} \otimes \I_9) \J \B \in \Rn{r \times r}$, and $\V^{v} \in \Rn{|\mathcal{T}| \times |\mathcal{T}|}$ being the diagonal $\mu$-weighed volume matrix $V^{v}_{ee} = \mu_e V_e $.

We turn our attention to the second term in this energy, $ \mu_e V_e tr(\F_e^T \R_{c(e)})$.
%fWe
With a similar treatment, we rewrite it in index notation,
\begin{align}
     \mu_e V_e tr(\F_e^T \R_{c(e)})  &= \mu_e V_e (\F_e^T \R_{c(e)})_{ii}, \\
     &=  \mu_e V_e  F_{ij} R_{c(e)ij} \nonumber,  \\
    &= \mu_e V_e  ({\mathcal{J}^{\B}}_{eijk}z_k) (P_{ec} R_{cij})  \nonumber ,\\
    &=  (\mu_e V_e  {\mathcal{J}^{\B}}_{eijk} P_{ec}) R_{cij} z_k ,
    \end{align}
where we have made use of a $\mathcal{|\mathcal{T}| \times |\mathcal{C}|}$  cluster selection matrix $\P$ to select out the tetrahedron inside each cluster.
\begin{align}
    P_{ec} = 
    \begin{cases}
    1 &  \mathrm{if} \; e \in \mathcal{T}(c),\\
    0 & \mathrm{otherwise}.
    \end{cases}.
\end{align}
With this selection matrix in hand, we can rewrite the second term going through the full summation,
\begin{align}
  -\sum_e^{|\mathcal{T}|}   \mu_e V_e tr(\F_e^T \R_{c(e)})  &=    -z_k   (  \sum_e^{|\mathcal{T}|}  \mu_e V_e  {\mathcal{J}^{\B}}_{eijk} P_{ec})R_{cij} , \nonumber \\
  &= z_k K^{v}_{ijkc} R_{cij}  \nonumber , \\
    &= \z^T ( \mathcal{K}^{v} : \R),
\end{align}
Where we've revealed the tensor $\mathcal{K}^{v} \in \Rn{r \times  3\times 3 \times |\mathcal{C}|}$, given by
\begin{align}
 \mathcal{K}^{v} = \mathrm{reshape}(  -\B^T \J^T ((\V^{v} \P)  \otimes \I_9), \{r, 3, 3,|\mathcal{C}|\}   ).
\end{align}

Abstracting this away further we arrive at 
\begin{align}
     -\sum_e^{|\mathcal{T}|}   \mu_e V_e tr(\F_e^T \R_{c(e)}) = \z^T \f^{v},
\end{align}

Finally, we can rewrite the reduced clustered passive muscle energy as a quadratic, 
\begin{align}
      \tilde{E}_v(\z) =  \frac{1}{2} \z^T \Q^{v} \z  + \z^T \f^{v} + \mathrm{const} , 
\end{align}

% \subsection{Expanding Active Muscle Energy $\tilde{E}_{\Phi}(\z)$}
\subsection{Expanding $\tilde{E}_{a}(\z, \a)$}
Starting with the actuation energy in \refeq{clustered-active-term}, we expand out the Frobenius norm:
% \begin{align}
%    \tilde{E}_{\Phi}(\z) &=  \sum_t^{|\mathcal{T}|} 
%    \frac{\gamma_t}{2}   || \F_t - \bOmega_{c(t)} \Y_t ||^2_F  \\
%      &= \sum_t^{|\mathcal{T}|}
%     \frac{\gamma_t}{2}  tr(\F_t^T \F_t) -  \gamma_t tr(\F_t^T \bOmega_{c(t)} \Y_t) + \mathrm{const}
% \end{align}
\begin{align}
   \tilde{E}_{a}(\z) &=  \sum_e^{|\mathcal{T}|} 
   \gamma_e V_e  || \F_e - \bOmega_{c(e)} \Y_e ||^2_F,  \\
     &= \sum_e^{|\mathcal{T}|}
    \gamma_e V_e tr(\F_e^T \F_e) -  2 \gamma_e V_e tr(\F_e^T \bOmega_{c(e)} \Y_e) + \mathrm{const}.
\end{align}
Just like the first section \refapp{Ev_derivation}, the first term amounts to the simple quadratic form 
\begin{align}
    \sum_e^{|\mathcal{T}|} \gamma_e V_e  tr(\F_t^T \F_t) = \z^T \Q^{a} \z  
\end{align}
with the actuation muscle Laplacian $\Q^{a} = \B^T \J^T (\V^{v} \otimes \I_{9})  \J \B \in \Rn{r \times r}$, where  $\V^{v} \in \Rn{|\mathcal{T}| \times |\mathcal{T}| }$ is the $\gamma$ weighed diagonal volume matrix such that $V^{v}_{ee} = V_e \gamma_e$.

For the second term we follow the same procedure as in the previous section, rewriting the integrand in index notation.
\begin{align}
   % (A B)_ij = A_ik B_kj
   % ((AB) C)_ij = (AB)_ik C_kj = (A_ic B_ck  C_kj
   % ((A^T B) C)_ij = (A_ci B_ck  C_kj)
   % (\F^T \bOmega \Y)_ij = (F_wi \Omega_wk  Y_kj)
   % tr(A) = A_ii
   % tr(\F^T \bOmega \Y) = (F_wi \Omega_wk  Y_ki)
   -\gamma_t tr(\F_t^T \bOmega_{c(t)} \Y_t) &=   - \gamma_t F_{twi} \Omega_{c(t)wk}  Y_{tki} \nonumber \\
   % F_twi = {\mathcal{J}^{\B}}_{twik} z_k
   % Y_tij = {\mathcal{J}^{\D}}_{tkil} \bar{a}_l
   &=   - \gamma_t ({\mathcal{J}^{\B}}_{twid} z_d)  (P_{tc}\Omega_{cwk})    ({\mathcal{J}^{\D}}_{tkil} \bar{a}_l)  \nonumber \\
   &=  \z_d (  -\gamma_t {\mathcal{J}^{\B}}_{twid}  {\mathcal{J}^{\D}}_{tkil}  P_{tc}  )    \Omega_{cwk}  \bar{a}_l \nonumber% \\
   % &=   \z_d  \mathcal{O}_{twkdlc}    \Omega_{cwk}  \bar{a}_l \nonumber 
\end{align}
% With $\mathcal{O} \in \Rn{|\mathcal{T}| \times 3 \times 3  \times m_s \times r \times |\mathcal{C}|}$. 
%
Plugging this into the sum allows us to express it in a our canonical form
\begin{align}
\begin{split}
   &\sum_e^{|\mathcal{T}|}    - \gamma_e V_e tr(\F_e^T \bOmega_{c(e)} \Y_e),    \\
   &=  z_d     ( \sum_e^{|\mathcal{T}|}    -\gamma_e V_e {\mathcal{J}^{\B}}_{ewid}  {\mathcal{J}^{\D}}_{ekil}  P_{ec}  )    \Omega_{cwk}  \bar{a}_l,   \\
   &=  z_d    \mathcal{O}_{wkdlc}    \Omega_{cwk}  \bar{a}_l ,  \\
    &=  \z^T ((\mathcal{O}\ \vdots\ \bOmega) : \bar{a}), \\
    &=  \z^t \f^{a},
    \end{split}
\end{align}
with $\mathcal{O} \in \Rn{  3 \times 3  \times r \times m_s \times |\mathcal{C}|}$. 
where $\f^{a} = ((\O\ \vdots\ \bOmega) : \bar{a})  \in \Rn{r}$.
We can finally rewrite our actuation energy as the quadratic,

\begin{align}
      % \tilde{E}^{a}(\z) =  \frac{1}{2} \z^T \Q^{a} \z  + \z^T \f^{a} + \mathrm{const}  
      \tilde{E}^{a}(\z) =  \z^T \Q^{a} \z  + 2 \z^T \f^{a} + \mathrm{const}  .
\end{align}
% \begin{align}
%     \sum_t^{|\mathcal{T}|} L_{tkl} z_k z_l =  z_k  S_t  L_{tkl}  z_l
% \end{align}

% \begin{align}
% \begin{split}
%     \tilde{E}_{\Psi}(\z) =  \min_{ \R_c}  \sum_t^{|\mathcal{T}|} 
%     \mu_t || \F_t - \R_{c(t)}  ||^2_F  \\
%      s.t. \; \forall \; c \in \mathcal{SO}(3) \forall{c} \in |\mathcal{C}|  .
%     \label{eq:clustered-passive-term}    
% \end{split}
% \end{align}

% \begin{align}
% \begin{split}
%         \tilde{E}_{\Phi}(\z, \L) =  \min_{ \bOmega_c}  \sum_t^{|\mathcal{T}|} 
%     \gamma_t || \F_t - \bOmega_{c(t)} \Y_t ||^2_F  \\
%      s.t. \; \forall \; c \in \mathcal{SO}(3) \forall{c} \in |\mathcal{C}|  .
%     \label{eq:clustered-active-term-2}    
% \end{split}
% \end{align}

\subsection{Expanding Inertial Energy $\tilde{E}_k(\z)$}

\begin{align}
    \tilde{E}_{k}(\z) =  \frac{1}{2h^2}|| \z - \q ||^2_{\N} &=   \frac{1}{2h^2}  \z^T \N \z  -  \frac{1}{h^2} \z^T \N \q  + \mathrm{const}.
\end{align}
Introducing $\Q^{k} = \frac{1}{h^2} \N  \in \Rn{r \times r}$, as well as $\f^{k} = - \frac{1}{h^2} \N \q \in \Rn{\r}$, we can write the kinetic energy simply as the quadratic energy, 
\begin{align}
E_{k}(\z) = \frac{1}{2} \z^T \Q^{k} \z  + \z^T \f^{k} + \mathrm{const}
\end{align}

\subsection{Expanding External Potential Energy  $\tilde{E}_f(\z)$}
This term is basically all expanded already, we only have to simplify it in our canoncial form,
\begin{align}
    E_f(\z) &= \z^T \B^T \M \g, \\
                &= \z^T \f^f,
\end{align}
with $\f^f =\B^T \M \g$.

\subsection{Putting it All Together}
Assembling all the terms of our energy that depend on $\z$ leaves us with the total quadratic energy, 
\begin{align}
   \tilde{E}(\z) = \frac{1}{2} \z^T (\Q^{a} +  \Q^{v} + \Q^{k}) \z  + \z^T (\f^{a} +  \f^{v} + \f^{k} + \f^f)   + \mathrm{const}.
\end{align}

Taking the minimizer of this quadratic energy finally reveals the system (without contact) we need to solve every timestep:
\begin{align}
    (\Q^{v} +  \Q^{a} + \Q^{k}) \z &= (\f^{a} +  \f^{v} + \f^{k})  \nonumber \\
    \Q \z = \f 
    \label{eq:appendix-global-step-no-contact}
\end{align}

Note that $\Q^v + \Q^a + \Q^k$ are constant matrices, and so the system factorization can be solved once and reused throughout the course of the entire optimization problem.

\begin{figure}
    \centering
    \includegraphics[width=\linewidth]{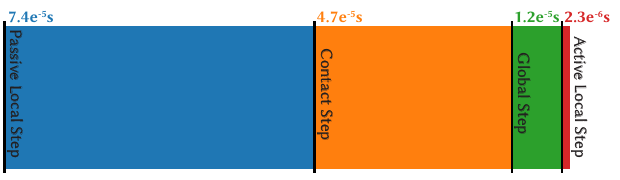}
    \caption{Timing breakdown for an average simulation step with our default 10 local-global iterations on the Arc de Triomphe character.}
    \label{fig:breakdown}
\end{figure}

\reffig{breakdown} shows our iteration step cost breakdown on the Arc de Triomphe character.

\section{Computing Contact Forces $\f_c$}
\label{app:contact}
To account for contact, we add an extra force term to the right hand side of \refeq{appendix-global-step-no-contact}, $\f_c \in \Rn{r}$,
\begin{align}
    \Q \z= \f + \f_c.
    \label{eq:appendix-global-step-final}
\end{align}

This extra force term $\f_c$ is computed every local-global iteration, before the global step to ensure $\z$ upholds the zero interpenetration, damping friction conditions,

\begin{align}
     \J^c|_N  \dot{\z} =  \0, 
     \J^c|_T \dot{\z} = u \J^c|_T\dot{\z}_{prev},
\end{align}
where $u$ is a velocity damping coefficient, $\J^c|_{N} \in \Rn{ |\mathcal{I}_c| \times r }$ is a contact Jacobian that maps from our reduced velocity  $\dot \z$ to per-contact point normal velocities. 
Likewise $\J^c|_{T} \in \Rn{ 2|\mathcal{I}_c| \times r }$ maps to tangential velocities.
The above two equations can be summarized by the constraint,
\begin{align}
    \J^c \dot{\z} = \vv,
\end{align}
where  ${\J}^c = [\J^c_N \; \J^c_T ]^T \in \Rn{ 3 | \mathcal{I}| \times r}$ is the full contact Jacobian and $\vv = [\boldsymbol{0} \;  u \J^c|_T\dot{\z}_{prev}]^T$.

Through finite differences, we rewrite our constraint on the velocity to act on positions DOFs $\z$,
\begin{align}
\begin{split}
         \J^c  \frac{\z - \z_{curr}}{h} =  \vv, \\
         \J^c  \z = h\vv  + \J^c  \z_{curr}  .
\end{split}
\end{align}

To compute an $\f$ that ensures $\z$ satisfies this condition, we plug in our global step solve $\z = \Q^{-1}(\f + \f_c)$ \refeq{appendix-global-step-final} into this constraint and obtain,

\begin{align}
\begin{split}
         \J^c    \Q^{-1} (\f + \f_c) &=   h\vv   + \J^c  \z_{curr},   \\
          \J^c    \Q^{-1}  \f_c &=   h\vv   + \J^c  \z_{curr} -  \J^c    \Q^{-1} \f , \\
          \L \f_c &= \b,
\end{split}
\label{eq:contact-constraint-on-z}
\end{align}

With $\L = \J^c    \Q^{-1}  $ and $\b =  h\vv   + \J^c  \z_{curr} -  \J^c    \Q^{-1} \f $.
The matrix $\Q^{-1} \in \Rn{r \times r}$ is a dense matrix, but it's also small and constant, and so the product $ \L =   \Q^{-1} \in \Rn{|\mathcal{I}| \times r}$ can be precomputed once and reused throughout the entire optimization procedure.
The system described by \refeq{contact-constraint-on-z} can either be over or underdetermined, depending on whether the number of contacting degrees of freedom $3|\mathcal{I}|$ is greater or less than the dimensionalilty of our simulation subspace $r$.

If it's overdetermined, we solve for $\f_c$ in a least squares sense,
\begin{align}
\begin{split}
    \f_c = \argmin_{\f_c} &||  \L \f_c -  \b  ||^2, \\
    &\Downarrow \\
    \L^T \L \f_c &= \L^T \b.
\end{split}
\end{align}

If it's underdetermined, we solve for the smallest $\f_c$ satisfying our constraint,
\begin{align}
\begin{split}
 \f_c =    \argmin_{\f_c}  ||  \f_c ||^2  \quad & \mathrm{s.t.}  \L \f_c = \b ,\\
   &  \Downarrow \\
    \f_c &= \L^T(\L \L^T)^{-1} \b .
\end{split}
\end{align}

\section{Tensor Precomputation for Clustering Rotations}
\label{app:tensor-precomp}

We start with the assumption that we have access to the simulation  Skinning Eigenmode Subspace $\B \in \Rn{3n \times r}$  \refeq{lbs_jacobian}, $\D \in \Rn{3n \times  (m_s + 1)}$ obtained from \refeq{motion-prior-spatial}, $\J \in \Rn{9 |T| \times 3n }$ the standard deformation Jacobian mapping matrix \cite{kim2022dynamicdeformables}.

% %
We construct the subspace deformation Jacobian mapping matrix $\J^B \in \Rn{9|T| \times r}$ and  $\J^D \in \Rn{9|T| \times (m_s + 1)}$:
\begin{align}
&\J^B = \J \B   &\J^D = \J \bar{\D} \nonumber
\end{align}

We reorder each of the entries in the form of a new 4-th order tensor $\mathcal{J}^B \in \Rn{|T| \times 3 \times 3 \times r}$ and
$\mathcal{J}^D \in \Rn{|T| \times 3 \times 3 \times  (m_s + 1)}$.
\begin{align}
 &\mathcal{J}^B = \mathrm{reshape}(\J^B, \{|T|, 3, 3, r\}),   \nonumber \\
 &\mathcal{J}^{\D} = \mathrm{reshape}(\J^D, \{|T|, 3, 3, m_s+1\}) \nonumber
\end{align}

% \begin{align}
%     \mathcal{J}^B &= \mathrm{reshape}(|T|, 3, 3, r),\\
%     \mathcal{J}^D &= \mathrm{reshape}(|T|, 3, 3, m_s).
% \end{align}

Which we can use to express the deformation Jacobian and the control target's deformation Jacobian matrices, $\F \in \Rn{|\mathcal{T}| \times 3 \times 3}$ and $\Y \in \Rn{|\mathcal{T}| \times 3 \times 3}$, using index notation, 
\begin{align}
    F_{eij} = \mathcal{J}^B_{eijk} z_k, & \quad 
    Y_{eij} = \mathcal{J}^D_{eijl} \bar{a}_l,
\end{align}
where a repeated index implies a component wise product along that dimension if the index is also present on the other side of the equation, or a standard dot product along that dimension if the index is not repeated on the other side of the equation.

We re-express  $\F_e \Y_e$, the matrix that appears inside the sum of \refeq{optimal-omega},
\begin{align}
     C_{eij} &=    F_{eik} Y_{ejk},  \nonumber \\
             &= (\mathcal{J}^B_{eiku} z_u) (\mathcal{J}^D_{ejkv} \bar{a}_v),  \nonumber \\
             &= (\mathcal{J}^B_{eiku} \mathcal{J}^D_{ejkv}) z_u \bar{a}_v, \nonumber  \\
             &= \mathcal{K}_{eijuv} z_u \bar{a}_v ,
\end{align}
revealing the constant tensor $\mathcal{K} \in \Rn{|\mathcal{T}| \times  3 \times 3 \times r \times  (m_s + 1) }$, which, when multiplied against the changing $\z$ and $\bar{\a}$, provides us with the per-tet matrix product of $\F_e \Y_e$ every timestep.

We then introduce a grouping matrix $\G \in \Rn{|\mathcal{C}| \times |\mathcal{T}|}$, a matrix that assembles for each cluster a  $\gamma$-weighed sum of its member tetrahedra's scalar quantities. 
\begin{align}
    G_{ce} = 
    \begin{cases}
    \gamma_e &  \mathrm{if} \; e \in \mathcal{T}(c)\\
    0 & \mathrm{otherwise}
    \end{cases}
\end{align}

This matrix allows us now to fully express the averaged covariance matrix $\H \in \Rn{|\mathcal{C}| \times 3 \times 3} $ in index notation, 
\begin{align}
         H_{cij} =  G_{ce}  C_{eij}
\end{align}
Finally, plugging in our expression for $C_{eij}$, we have
\begin{align}
        H_{cij} &= G_{ce} \mathcal{K}_{eijuv} z_u \bar{a}_v \nonumber \\
        &= \mathcal{H}_{cijuv} z_u \bar{a}_v
\end{align}

Where we finally expose the constant tensor $ \mathcal{H}_{cijuv} $, which multiplies $\z$ and $\bar{\a}$ and provides the covariance matrices for all clusters $\H$.

\section{Optimal Clustered Rotations}
\label{app:optimal-clustered-rotations}
% We want to show that the active clustered best-fit rotation $\tilde{\Omega}_c$. 
We want to derive the optimal rotation $\bOmega_c$ for each cluster $c$.
%
% We will drop the dependence of $\F_e(\z)$ on and $\Y_e(\a)$ for clarity. 
% \begin{align}
%      \tilde{\Omega}_c &=  \argmin_{\tilde{\Omega}_{c} \in \mathcal{SO}(3)}\sum_t^{|T(c)|} \gamma_t || \F_t - \tilde{\Omega}_{c} \Y_{t} ||_F^2 &\quad \forall c \in |\C| 
% \end{align}
%
Starting from \refeq{clustered-active-term}, computing the optimal rotation $\bOmega_c$ for each $c$ amounts to the following optimization problem
\begin{align}
     \bOmega_c &=  \argmin_{\bOmega_{c} \in \mathcal{SO}(3)}\sum_e^{|T(c)|} \gamma_t V_e || \F_e - {\bOmega}_{c} \Y_{e} ||_F^2 
\end{align}
As presented in \refeq{optimal_rotations}, we will show that the optimal solution is given by the polar decomposition:
%
% \begin{align}
%     \tilde{\Omega}_c &= \f_{p}(\K_c)  = \mathrm{polar} \left( \sum_t^{|T(c)|} \gamma_t \F_t \Y_t^T \right)  \; .
% \end{align}
%
\begin{align}
    \bOmega_c  = \mathrm{polar} \left( \sum_e^{|\mathcal{T}(c)|} \gamma_e V_e \F_e \Y_e^T \right).
    \label{eq:optimal_rotations_appendix}
\end{align}
The same derivation for the active clustered rotation $\bOmega_c$ can be extended to the optimal rotation $\R$ for the passive term $E_v$ by setting $\Y_e= \boldsymbol{I}_{3 \times 3}$ for every tetrahedral element $e$.

\paragraph{Proof} 
We start by expanding each term in the summation of \refeq{optimal_rotations_appendix} as
\begin{align}
    % \Psi(\z) &= 
    &\gamma_e V_e || \F_e - {\bOmega}_{c} \Y_{e} ||_F^2  \\
    &= \gamma_e V_e  tr\left(( \F_e - {\bOmega}_{c} \Y_{e})^T ( \F_e - {\bOmega}_{c} \Y_{e})\right) \\
    &= \gamma_e V_e  tr\left( \F_e^T \F_e + \Y_e^T \Y_e -  \F_e^T  {\bOmega}_{c}   \Y_e -  \Y_e^T  {\bOmega}_{c}^T   F_e \right) \\
    &= -\gamma_e V_e  tr\left( 2\F_e^T  {\bOmega}_{c}   \Y_e \right) + \mathrm{const} \\
        &= -\gamma_e V_e  tr\left( 2 {\bOmega}_{c}   \Y_e \F_e^T \right) + \mathrm{const}
\end{align}
Above, in the second equality we made use of the identity $||\A||^2_F = tr(\A^T \A)$.
In the third equality we expanded the multiplication, and used the orthogonality of rotation matrices ${\bOmega}_{c}^T {\bOmega}_{c} = \I$. 
In the fourth equality we dropped terms that do not depend on the variable we are optimizing over and placed them inside the constant $c$. 
In the fifth and final equality we have made use of the trace's invariance under permutation of its input.
We've also used the identity that the trace is invariant under transposes $tr(A + A^T) = tr(2A)$.

With the above result, we return to our full optimization problem (\refeq{optimal_rotations_appendix}) and rewrite it as:
\begin{align}
     {\bOmega}_c &=  \argmin_{{\bOmega}_{c} \in \mathcal{SO}(3)}\sum_e^{|T(c)|} - \gamma_e V_e  2tr\left( {\bOmega}_{c}   \Y_e \F_e^T  \right) + \mathrm{const} \\ % &\quad \forall c \in |\C|  \\
      &=  \argmax_{{\bOmega}_{c} \in \mathcal{SO}(3)}\sum_e^{|T(c)|} \gamma_e V_e  tr\left( {\bOmega}_{c}   \Y_e \F_e^T  \right) \\ % &\quad \forall c \in |\C|  \\
   &=  \argmax_{{\bOmega}_{c} \in \mathcal{SO}(3)}   tr\left( \sum_e^{|T(c)|} \gamma_e V_e  {\bOmega}_{c}   \Y_e \F_e^T  \right) \\ % &\quad \forall c \in |\C| \\
   &=  \argmax_{{\bOmega}_{c} \in \mathcal{SO}(3)}   tr\left( {\bOmega}_{c} 
 \sum_e^{|T(c)|}  \gamma_e V_e \F_e^T \Y_e  \right)  \\
   &= \argmax_{{\bOmega}_{c} \in \mathcal{SO}(3)}   tr\left( {\bOmega}_{c} \H_c \right) 
\end{align}
In the second equality we have dropped the $\mathrm{const}$ and switched to a maximization instead of a minimization, while also dropping the factor of 2 from the optimization. 
In the third equality we have made use of the linearity of the trace operator and the sum operator, allowing them to commute.
In the fourth equality, we finally pull out the per-cluster rotation ${\bOmega}_{c}$ from the sum, as all tets in $|T(c)|$ share the same rotation matrix by definition. 
This final form is an instance of the \emph{Orthogonal Procrustes} problem, where the target matrix we are trying to match is $\H_c$.
Based on the proof in \citet{schoenemann1964solution, enwiki:1190317617}, the optimal solution to  the Orthogonal Procrustes problem can be obtained from the polar decomposition of $\H_c$. 
\begin{align}
    {\bOmega}_c  = \mathrm{polar}(\H_c) = \mathrm{polar} \left( \sum_e^{|\mathcal{T}(c)|} \gamma_e V_e \F_e \Y_e^T \right) \label{eq:optimal-omega}
\end{align}

\qedsymbol

\end{document}